\documentclass[prl,aps,showpacs,floatfix,floats,endfloats,nobalancelastpage,preprint]{revtex4}
\usepackage{graphicx,bm,amsmath}    

\begin{document}
\title{Semiclassical path integral approach on spin relaxations in narrow wires}
\author{Cheng-Hung Chang$^{1,2,3}$, Jengjan Tsai$^1$, Hui-Fen Lo$^4$, A.G. Mal'shukov$^5$
}
\date{\today}
\affiliation{
$^1$ Institute of Physics, National Chiao Tung university, Hsinchu 300, Taiwan \\
$^2$ National Center for Theoretical Sciences, Physics Division, Hsinchu 300, Taiwan \\
$^3$ Institute of Mathematical Modeling and Scientific Computing,
National Chiao Tung university, Hsinchu 300, Taiwan \\
$^4$ Department of Physics, National Tsing-Hua University, Hsinchu 300, Taiwan \\
$^5$ Institute of Spectroscopy, Russian Moscow Academy of Science,
142190 Troitsk, Moscow oblast, Russia }



\begin{abstract}
The semiclassical path integral (SPI) method has been applied for
studying spin relaxation in a narrow 2D strip with the Rashba
spin-orbit interaction. Our numerical calculations show a good
agreement with the experimental data, although some features of
experimental results are not clear yet. We also calculated the
relaxation of a uniform spin-density distribution in the ballistic
regime of very narrow wires. With the decreasing wire width, the
spin polarization exhibits a transition from the exponential decay
to the oscillatory Bessel-like relaxation. The SPI method has been
also employed to calculate the relaxation of the particularly
long-lived helix mode. A good agreement has been found with
calculations based on the diffusion theory.

\end{abstract}



\maketitle

\section{I Introduction}

The spin relaxation rate is an important spin transport parameter.
Recent calculations and measurements of this parameter in
semiconductor systems have to a great extent been motivated  by
numerous ideas of spintronic applications. \cite{reviewspintronics}
In view of these applications, as well as from the fundamental point
of view, one of the most interesting problem is the spin relaxation
in quantum dots (QD) and quantum wires (QW). In zincblende
semiconductors at low temperatures the spin lifetime is mainly
determined by the D'yakonov-Perel' \cite{DP} mechanism associated
with spin-orbit effects. In systems with restricted dimensions this
relaxation mechanism is strongly suppressed, as has been calculated
in the case of QD \cite{Chang1} and QW
\cite{Malshukov,Kiselev,Schwab}. The physics of such a suppression
in QW became clear from the analytical solution of the diffusion
equation for nonuniform spin distributions confined in a
wire.\cite{Malshukov} Surprisingly the suppression starts when the
width $w$ of a wire becomes less than the characteristic length
$L_{so}$ of the spin orbit interaction. In typical zincblende
semiconductor systems it varies from several thousands \AA~  up to
several microns and can be much larger than the electron mean free
path $l$. Hence, such a spin lifetime enhancement can not be
considered as a manifestation of the motional narrowing effect when
a restricted geometry of the system imposes the upper limit on the
mean free path. Indeed, recent measurements \cite{Awschalom} have
demonstrated that the spin lifetime $\tau_s$ starts to increase
already at $w \gtrsim 10\,l$. On the other hand, the observed
slowdown of the spin relaxation appears to be not so strong, as
expected from the theory. To understand such a behavior, one has to
take into account that in experiments \cite{Awschalom} the measured
parameter is the relaxation time of a particular spatial spin
distribution, rather than of an individual electron spin. At the
same time, as shown in Ref. \onlinecite{Malshukov} only two kinds of
spin distributions have very long lifetimes in narrow 2D wires. The
first one corresponds to a polarization which is homogeneous along
the wire with spins oriented in the plane of a 2D electron gas
(2DEG) and perpendicular to the wire axis. The second distribution
is a nonuniform helix mode with the wavelength determined by
$L_{so}$. As it will be pointed out below, none of these
distributions have been excited by an incident light beam in the
experiment \cite{Awschalom}.

In order to interpret experimental data we will analyze relaxation
of various spin distributions. We will study diffusive, as well as
ballistic regimes of electron motion in the wire. The path integral
method previously applied to QD \cite{Chang1} will be employed to
calculate the spin relaxation in a wide parameter range, including
the ballistic regime $w \lesssim l$, and at time intervals less than
the electron momentum relaxation time.

The article is organized by the following way: sec.II gives an
introduction to the path integral method; in sec. III the spin
relaxation of a homogeneous spin distribution is calculated and a
comparison with the experiment is given; in sec. IV some analytical
results are presented useful for understanding the spin relaxation
behavior in the ballistic range; sec. V is devoted to an analysis of
long-lived helix spin distributions, beyond the diffusion theory of
Ref. \onlinecite{Malshukov}. The conclusion is presented in sec. VI.

\section{II Semiclassical path integral approach}

The semiclassical path integral (SPI) formalism has been used to
study the  spin relaxation and spin current transmission in systems
with the Rashba spin-orbit interaction (SOI)
\cite{Chang1,Chang2,Malshring}. The Hamiltonian of these systems can
be divided into two parts
\begin{equation}\label{eq2_1}
H=H_0+H_{R},
\end{equation}
where $H_0$ contains the kinetic and potential energies of an
electron in a 2DEG. The second part $H_R=\alpha\,(\bm{p}\times
\bm{\sigma})\cdot\bm{z}$ represents the SOI, where $\alpha$ is the
spin-orbit coupling constant, $\bm{p}$ denotes the electron
momentum, $\bm{\sigma}$ stand for Pauli matrices, and $\bm{z}$ is
the unit vector perpendicular to the 2D sample. Within the SPI
method the spin-orbit interaction $H_{R}$ gives rise to spin
precession of a particle moving along a classical trajectory. A
characteristic length determining this precession is given by
$L_{so}=\frac{\hbar}{\alpha m^{*}}$, where $m^*$ is the effective
mass of the particle. In real semiconductor materials, the energy
ratio $H_{R}/H_0$ can reach $1/10$, like in the InSb
sample.\cite{Chen} But even for such a ratio, $H_R$ is still small
compared with $H_0$. In such systems whose characteristic size, or
the electron mean free path, are smaller than $L_{so}$, the electron
dynamics is not affected strongly by its spin dynamics, so that in
the leading approximation classical trajectories are determined by
$H_0$.

For a free electron moving along a straight trajectory $\gamma$ of
length $l$, the dynamics of its spin state is governed by the
evolution operator $U$ in the path integral formalism \cite{Chang1}
\begin{equation}\label{eq2_2}
U=\exp\left[-\frac{i}{\hbar}\int_{\gamma}H_{R}(t)\,dt\right]
=\exp\left[-i\frac{l}{L_{so}}\bm{b}\cdot\bm{\sigma}\right],
\end{equation}
where $\bm{b}=\bm{z}\times\bm{p}/|\bm{p}|$. This operator represents
simply the spin rotation. Note that the semiclassical approximation
has been employed in Eq. (\ref{eq2_2}), since only the classical
paths were taken into account in the integral. \cite{Chang1} This
approximation is valid for systems whose size is much larger than
the de Broglie wavelength of electrons. The term `semiclassical'
here is referred to the semiclassical (saddle point) approximation
taken in the spin evolution operator in Eq. (2). Though this
terminology is used, the following analysis within the SPI method is
classical one. This method is equivalent to the classical Boltzmann
equation modified to take into account spin dynamics. Such an
equation was used already in Ref. \cite{DP} and many times since.

If an electron collides with impurities or boundaries $n_\gamma-1$
times, its trajectory $\gamma$ will consist of $n_{\gamma}$ straight
segments
$${\gamma}={\gamma}_{n_{\gamma}}+\cdots+{\gamma_{j}}+\cdots+{\gamma_{2}}+{\gamma_{1}},$$
The corresponding spin evolution operator $U_{\gamma}$ becomes a
product
\begin{equation}\label{eq2_3}
U_{\gamma}=U_{\gamma_{n_{\gamma}}}\cdots U_{\gamma_{j}}\cdots
U_{\gamma_{2}}U_{\gamma_{1}},
\end{equation}
where the individual operators
\begin{eqnarray}\label{eq2_4}
U_{\gamma_{j}}&=&\exp\left[-i\frac{l_{j}}{L_{so}}\bm{b}_{j}\cdot\bm{\sigma}\right]
\nonumber \\
&=&{\bf 1}\cos\left(\frac{l_{j}}{L_{so}}\right)-i
(\bm{b}_{j}\cdot\bm{\sigma})\sin\left(\frac{l_{j}}{L_{so}}\right)
\end{eqnarray}
along different straight segments do not commute with each other.
Each trajectory $\gamma$ at the time $t$ is uniquely determined by
the particle initial coordinate and momentum. Since in the
semiclassical approximation the particle momentum is equal to the
Fermi momentum $p_F$, the trajectory will depend on its initial
angle, while its total length is simply $tv_F$. At the time $t$ the
spin of a particle moving along the trajectory $\gamma$ is
determined by $\mathbf{s}^{(\gamma)}(t,\sigma_0)= \langle
U_{\gamma}^{-1}\bm{\sigma}U_{\gamma}\rangle_{\sigma_0}$, where
angular brackets denote averaging over the initial spin state
$\sigma_0$. Using this expression one may calculate the evolution of
the spin state in systems of various geometries, with or without
elastic impurity scatterers, as shown by several examples in Ref.
\onlinecite{Chang1}. In order to determine the evolution of a given
particle distribution, one must perform a statistical average of the
above expression over initial trajectory coordinates and angles,
which we will denote by the trajectory label $\gamma$. Below, we
will assume that particle initial positions and momentum directions
are uniformly distributed. In this case, given a small area $D$ of
2DEG, the $z$ projection of the spin polarization $P_z(t)$ within
this area at time $t$ is determined by the average
\begin{equation}\label{eq2_5}
P_z(t)=\frac{1}{n_{(t,D)}}\sum_{\gamma(D)}
s_z^{(\gamma)}(t,\sigma_0),
\end{equation}
where $n_{(t,D)}=\sum_{\gamma(D)}$ is the number of trajectories
reaching the area $D$ at time $t$, with $s_z^{(\gamma)}(t,\sigma_0)$
denoting the $z$-component of the electron spin. According to the
above definition, $s_z^{(\gamma)}(t,\sigma_0)$ varies within
$[-1,1]$. Hence, the maximum value of $|P_z(t)|$ is $1$, which
corresponds to all electrons in $D$ being aligned in $z$-direction.

To apply the SPI method numerically, a large number of electrons are
initially randomly distributed in the channel with uniform or helix
spin configurations as explained below. Each electron moves straight
before collision with impurities. The distance between two
collisions follows the well known exponential distribution of free
paths. In the following, we assume that the channels have smooth
boundaries on which the electron reflection is specular. In the
diffusive regime (as in the experimental sample in Ref.
\cite{Awschalom}), the spin relaxation behavior under this
assumption is the same as in the case of non-smooth boundaries,
because, even when electron trajectories are not randomized by the
smooth boundary, they will be immediately randomized by the
impurities near the boundaries. In the ballistic regime, the
relaxation behaviors in systems with smooth and non-smooth
boundaries are different. Here we focus on the simple example of
specular reflection. Once the boundary roughness of a ballistic
sample is known, the extension to the non-specular case is
straightforward.

\section{III Relaxation of uniform spin modes}

The spin relaxation times obtained in the experiments of Ref.
\cite{Awschalom} were measured in a 2D n-InGaAs channel of the
length $L = 200\,\mu$m and the width $w = 0.42\sim 20\,\mu$m. The
SOI in the sample is dominated by the Rashba coupling. In the
notation of Ref. \onlinecite{Awschalom} it corresponds to the
characteristic length $l_{SP}\simeq 1\,\mu$m, which is related to
the above defined spin rotation length $L_{so}$ as
$L_{so}=2\,l_{SP}$.

The sample is characterized by the electron mean free path $l =
0.28\,\mu$m, the momentum scattering time $\tau_M=0.76$ ps, and  its
Fermi velocity can accordingly be estimated as $v_F
=0.28\,\mu$m$\div 0.76$ ps $\approx 0.37\,\mu$m/ps. For the carrier
concentrations $n_s=5.4 \sim 7.0\times 10^{11}$ cm$^{-2}$ used in
Ref. \cite{Awschalom}, the de Broglie wavelength
$\lambda_f=\sqrt{2\pi/n_s}$ of electrons in the 2DEG is around
$30\sim 34$ nm. The sample was patterned along various
crystallographic directions and electron spins have been optically
oriented parallel to the growth direction $[0,0,1]$. The relaxation
times $\tau_s$ measured in Ref. \cite{Awschalom} are replotted by
the circles and the stars connected by the blue and the green curves
in Fig. 1.

\begin{figure}[htbp!]
\center{\includegraphics[width=15cm]{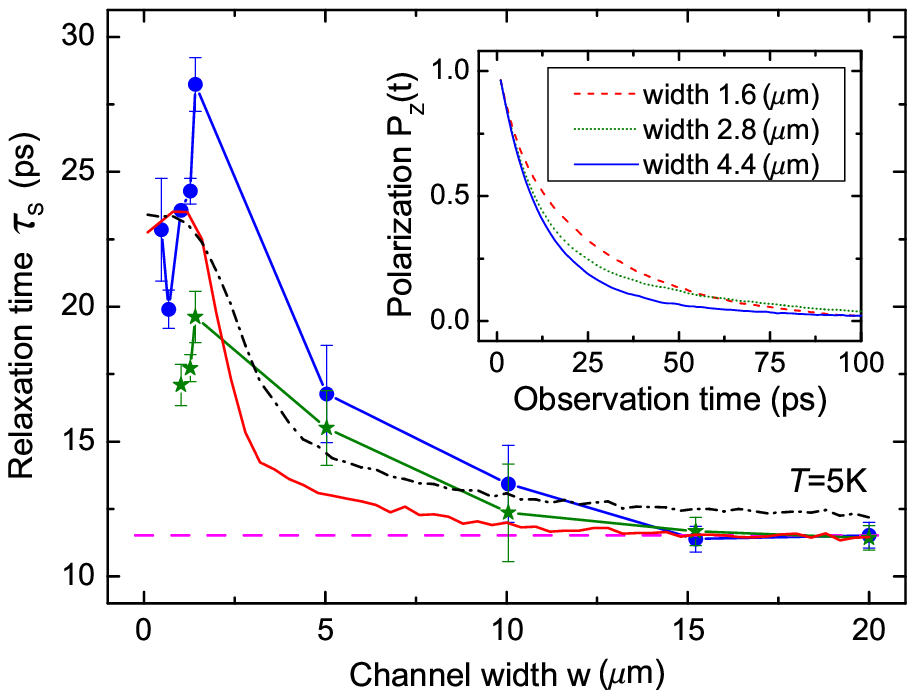}}
\caption{
(Color online) The spin relaxation times $\tau_s$'s of $[1,0,0]$
sample (circle) and $[1,1,0]$ sample (star) versus channel width $w$
are taken from the experiments in Ref. \cite{Awschalom}. The
$\tau_s$ calculated by the SPI method are extracted from the
polarization curve $P_z(t)$, fitted by Eq. (\ref{eq3_1}) with free
parameters $A$ and $c$ (red solid curve) and with fixed parameters
$A=1$ and $c=0$ (black dash-dotted curve). The experimental $\tau_s$
saturates at $11.5\,\mu$m (pink dashed straight line) for large $w$,
the same as the analytically estimated value for
$L_{so}=2.19\,\mu$m. The inset demonstrates three examples of
$P_z(t)$ for channel width $w=1.6\,\mu$m, $2.8\,\mu$m, and
$4.4\,\mu$m.}
\end{figure}

Since the width range $0.4\,\mu$m $\leq w\leq 20\,\mu$m used in our
calculations is much larger than the de Broglie wavelength
$\lambda_f$, the quantum effects are negligible and the validity of
the SPI approach is justified. With the above experimental
parameters, the SPI calculations are represented in Fig. 1, where
the inset shows the relaxation curves $P_z(t)$ for three channels of
different widths. All electron spins were initially aligned in
$z$-direction. The relaxation time $\tau_s$ can be determined by a
fitting of these $P_z(t)$ curves with the exponential function
\begin{equation}\label{eq3_1}
P_z(t)=A\exp(-t/\tau_s)+c.
\end{equation}
For example, the (red) solid curve in Fig. 1 represents the
relaxation time of $1.2\times 10^7$ electrons in channels of
different widths $w$'s. A comparison with the experimental data
(circles and stars) leads to following conclusions:
\begin{itemize}
\item[(i)] At large widths ($w>15$ $\mu$m), the electron spin can be regarded
as relaxing in bulk systems. In the experiments in Ref.
\onlinecite{Awschalom}, $l_{SP}$ was estimated to be $1.0\pm
0.1\,\mu$m, corresponding to $L_{so}=2.0\pm 0.2\,\mu$m. This
experimental uncertainty results in $\tau_s=9.7\pm 2.1$ ps, when
calculated by the SPI method. However, each $\tau_s$ obtained from
the SPI method agrees very well with that determined by the
analytical expression of DP relaxation $\tau_s=L_{so}^2/(4v_F\,l)$
for boundless systems. Thus, if the experimental samples are
governed by pure Rashba Hamiltonian, as in our calculation, these
samples most likely have $L_{so}=2.19\,\mu$m. This value is used in
our SPI simulations to obtain the red and black curves in Fig. 1.

\item[(ii)] For intermediate widths ($1.4\,\mu$m$<w<15\,\mu$m),
there is no an analytical expression for $\tau_s$ to compare with.
The SPI result deviates slightly from the experimentally measured
$\tau_s$. The maximum deviation is around 3 ps for $[1,1,0]$ sample
and 4 ps for $[1,0,0]$ sample at $w=5\,\mu$m. The calculated
$\tau_s$ is closer to the $\tau_s$ of the $[1,1,0]$ sample.

\item[(iii)] For small widths ($w<1.4\,\mu$m), the experimentally measured
$\tau_s$ saturates at $28$ ps for $[0,0,1]$ sample and $20$ ps for
$[0,1,1]$ sample. It is assumed in Ref. \onlinecite{Awschalom} that
this saturation might be related to other mechanisms, like the bulk
inversion asymmetry. However, the calculated $\tau_s$ in Fig. 1 is
also bounded by a maximum value around $24$ ps, although in our
calculation only the Rashba Hamiltonian was considered, without any
additional mechanisms involved.
\end{itemize}

\begin{figure}[htbp!]
\center{\includegraphics[width=15cm]{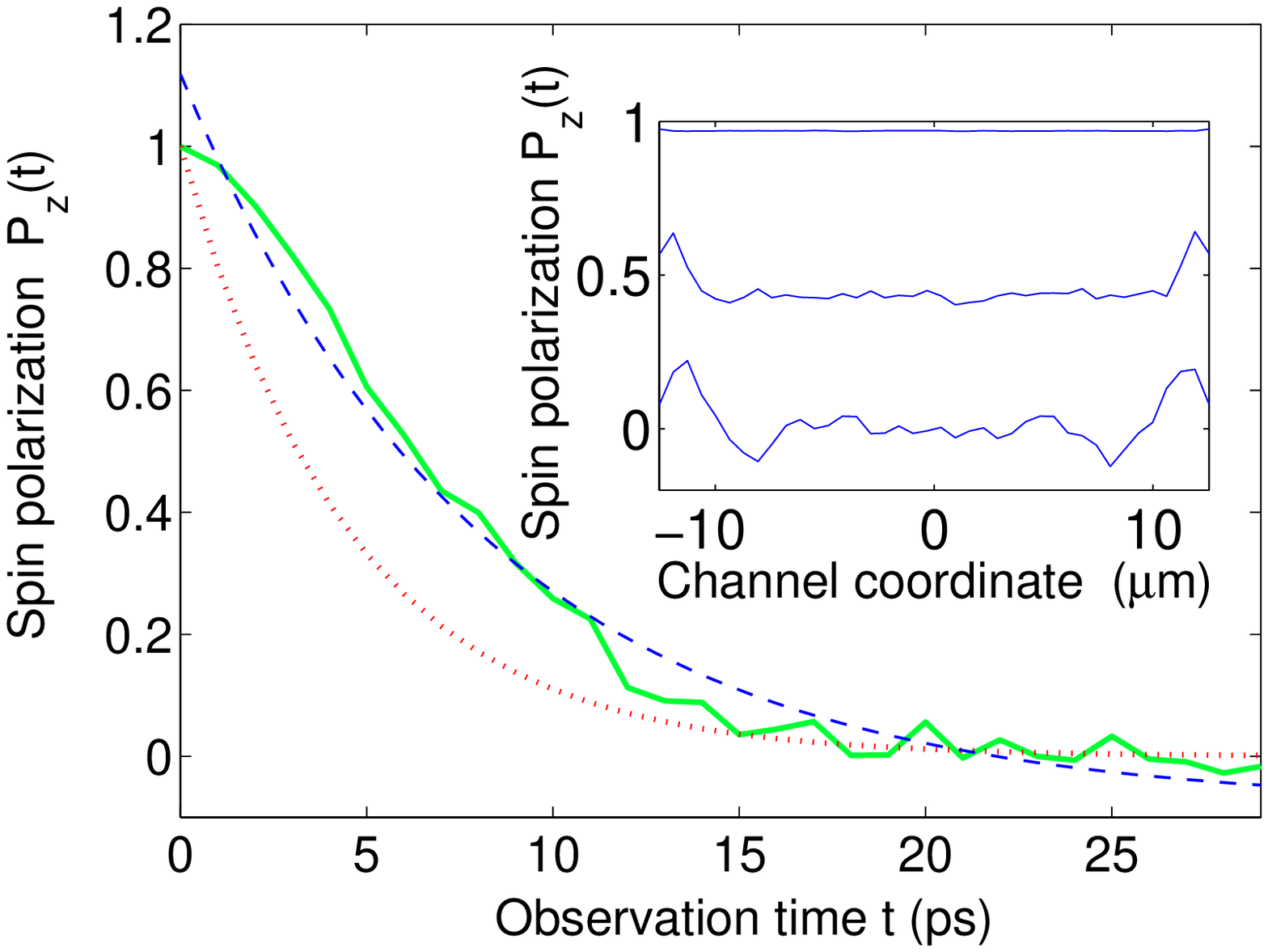}}
\caption{
(Color online) In the inset, a spin configuration in a channel of
the length $8\pi\;\mu$m relaxes to zero, depicted at three different
times. These configurations are spatially uniform up to the ripples
at two ends caused by boundary effect. Recording the polarization
$P_z(t)$ at the middle point of the channel gives the relaxation
curve (green solid thick) in the main plot. This curve can be fitted
by the exponential function in Eq. (\ref{eq3_1}) with
$[\tau_s,A,c]=[7.127,\,1.274,\,-0.055]$ (blue dashed curve) and
$[\tau_s,A,c]=[4.323,\,1,\,0]$ (red dotted curve). For $t$ close to
zero, most electrons have not been reflected by impurities or
boundaries. In this range $P_z(t)$ does not behave as an exponential
function. Later, after most of the electrons and their spins have
been randomized by impurities or boundaries, $P_z(t)$ became more
exponential-like. The physical parameters used are $[w, L_{so}, v_F,
l] =[0.1\,\mu{\rm m}, 2\,\mu{\rm m}, 0.37\,\mu{\rm m/ps},
0.3\,\mu{\rm m}]$ with $6\times 10^4$ electrons.}
\end{figure}

An important factor affecting the interpretation of the experimental
data is how to determine the relaxation time $\tau_s$ from the
function $P_z(t)$. The solid curve in Fig. 2 is an example of
$P_z(t)$ in a channel with $w=0.1\,\mu$m and $l=0.3\,\mu$m. At first
sight it looks like an exponential function to be fitted with a
relaxation time $\tau_s$ in Eq. (\ref{eq3_1}). But a closer look
shows that it is not a pure exponential function. Indeed, if we
gradually increase $l$ by reducing the number of impurities in the
channel, the monotonically decreasing $P_z(t)$ in Fig. 2 will
transform to an oscillatory function. In the extreme case of an
infinitely thin impurity-free channel, we shall prove in the next
section that the evolution of $P_z(t)$ will follow the Bessel
function
\begin{equation}\label{eq3_2}
    P_z(t)=J_0\left(\frac{2v_F\,t}{L_{so}}\right).
\end{equation}
This analytical formula is depicted by the smooth (red) dashed curve
in Fig. 3. A corresponding result of a numerical SPI simulation is
plotted as a (green) rugged solid curve in the same figure.

\begin{figure}[htbp!]
\center{\includegraphics[width=15cm]{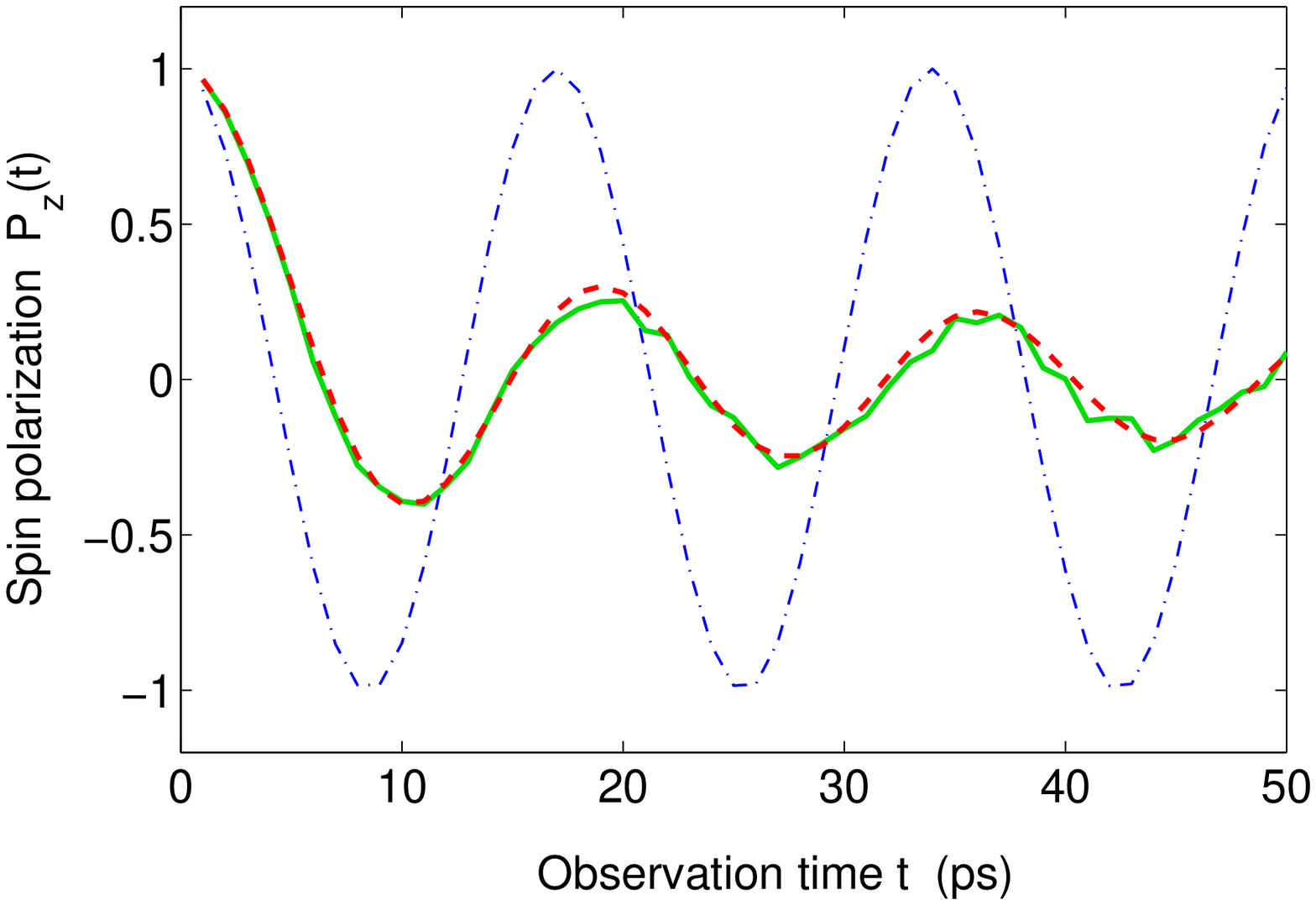}}
\caption{
(Color online) In a 1D channel without impurities, a spin
polarization $P_z(t)$ behaves like a sinusoidal function Eq.
(\ref{eq4_1}) (blue dash-dotted curve). In an infinitely thin
channel without impurities, $P_z(t)$ behaves like a Bessel function
Eq. (\ref{eq3_2}) (smooth red dashed curve), which agrees with the
numerically obtained $P_z(t)$ simulated by $5\times 10^4$ electrons
(rugged green solid curve). The physical parameters are $[w, L_{so},
v_F, l] =[0.1\,\mu{\rm m}, 2\,\mu{\rm m}, 0.37\,\mu{\rm m/ps},
10^4\,\mu{\rm m}]$.}
\end{figure}

During transition from the diffusive to the ballistic regime,
$P_z(t)$ will undergo a crossover from an exponential to a Bessel
function. In principle, it is meaningless to use an exponential
function to extract $\tau_s$ from such a crossover function,
especially when it is far from an exponential behavior. But if one
would like to carry out this procedure, the so obtained $\tau_s$
will depend on the choice of the parameters $A$ and $c$ in Eq.
(\ref{eq3_1}):
\begin{itemize}
\item[(A)] If $A=1$ is chosen, Eq. (\ref{eq3_1}) can precisely fit
the real initial polarization $P_z(t)=1$ at $t=0$ (red dotted curve
in Fig. 2).

If $A\neq 1$, Eq. (\ref{eq3_1}) can provide a better fitting to
$P_z(t)$ in a wider range of times at $t>0$ (blue dashed curve in
Fig. 2). On this reason, such a choice of A seems to be more
appropriate.

\item[(B)] Further, if $c\neq 0$ the fitted
values of $c$ and $\tau_s$ will be strongly dependent on the
observation time cutoff. The reason is that usually the tail of
$P_z(t)$ is oscillating, if the system within the considered range
of times is not in the diffusive regime. The closer the system to
the ballistic regime, the larger is the oscillation amplitude. The
Bessel function in Eq. (\ref{eq3_2}) for `pure' ballistic regime has
the largest amplitude. If the non-oscillating Eq. (\ref{eq3_1}) is
used to fit an oscillating Eq. (\ref{eq3_2}) truncated at some
cutoff, the fitted $\tau_s$ and $c$ will depend on the cutoff. The
corresponding uncertainty of $\tau_s$ will decrease with an
increasing observation time.
\end{itemize}

In the experiments \cite{Awschalom}, the width of the channel varies
between $w\approx 1.5\,l$ and $70\,l$. Since at the smallest $w$ the
system is not far from the ballistic regime, the difference between
$P_z(t)$ and the exponential function should be observable. Indeed,
the value of $\tau_s$  fitted by Eq. (\ref{eq3_1}) with $A=1$ and
$c=0$ (black dash-dotted curve in Fig. 1) is somewhat distinct from
$\tau_s$ at $A\neq 1$ and $c\neq 0$ (red solid curve in Fig. 1).
Since our observation time is sufficiently long, the fitted value of
$c$ is close to zero. A disagreement produced by different fitting
procedures will become more remarkable when the system approaches
the ballistic regime with strongly oscillating $P_z(t)$. Hence, when
comparing $\tau_s$'s obtained by different research groups, it is
important to know the whole set of the fitting parameters ($A$, $c$,
and observation time). Even when the same $P_z(t)$ curve is
considered, the reported $\tau_s$'s could be different. One more
problem with the fitting procedure is that even in the diffusive
regime the evolution of the spin polarization not necessarily
follows the exponential behavior with a single relaxation time. For
example, a homogeneous $P_z$ distribution is not an eigenstate of
the diffusion equation in a 2D channel. Therefore, as shown in
Ref.~\onlinecite{Schwab}, edge states can contribute to the $P_z(t)$
evolution with the relaxation time different from that of the bulk
eigenstate. The weight of edge states increases with decreasing $w$.

In regime (ii), the experimental data deviate slightly from the SPI
calculations with a maximum difference $\tau_s\approx 3\sim4$ ps at
$w\approx 5\,\mu$m. This discrepancy is too large to be attributed
to different fitting procedures. One of the explanations for such a
behavior might be a specific role of long-lived edge states. The
lifetime of such modes depends on the boundary conditions
\cite{Schwab}. Our SPI calculations assumed a specular reflection of
electrons from hard wall boundaries of the wire. Probably, the
experimental situation in Ref.\onlinecite{Awschalom} corresponds to
other boundary conditions which give rise to the edge states with
larger $\tau_s$. This problem requires a more thorough analysis.

In regime (iii), the relaxation time goes to a finite value at $w
\rightarrow 0$ in both experimental and SPI calculated plots in Fig.
1. For a homogeneous spin distribution along the channel, the
diffusion theory \cite{Malshukov} also predicts a saturation of
$\tau_s$ at $w \rightarrow 0$. The saturated value should be twice
of the bulk DP spin relaxation time. With the experimental bulk
value $\tau_s$=11.5 ps, one expects $\tau_s$=22.8 ps at $w =0$.
Experimental and SPI curves at Fig.1 are not far from this value,
although the diffusion approximation fails at $w \simeq l$. At the
same time, one should not forget that in a narrow channel the time
evolution of the spin polarization strongly deviates from the
exponential function. On this reason, in regime (iii) $\tau_s$ can
not be a representative parameter to describe the spin relaxation.

\section{IV Bessel relaxations in ballistic channels}

Depending on the ratio between the channel width and the electron
wavelength, one encounters two limiting cases. If the wire carries
only one propagating channel, we have effectively a 1D situation. In
the opposite limit, if the width of the wire is much larger than the
electron wavelength, semiclassical electrons are able to move in
both $x$- and $y$-directions. Therefore, the system is
two-dimensional, even if geometrically the channel is narrow, with
$w$ much less than other characteristic lengths, such as $L_{so}$
and $l$. Below, we will consider the evolution of electron
polarization in the ballistic regime for these two limiting cases.

\subsection{1D ballistic channels}

Let us consider a 1D impurity-free channel where at $t=0$ spins of
all electrons are aligned in $z$-direction. Since impurities and the
electron-electron interaction are absent, electrons can only move in
$+x$ or $-x$ directions along the channel axis with a constant
velocity. The spins of all these electrons will rotate
simultaneously along different geodesics connecting the north and
south poles on the spin sphere. A $2\pi$ spin rotation takes place
when an electron passes a distance $\pi L_{so}$ during a time period
$\pi L_{so}/v_F$. Therefore, the angular frequency of this rotation
is $2\,v_F/L_{so}$, the same for all spins. The spin polarization at
any place in the channel will then evolve according to
\begin{equation}\label{eq4_1}
    P_z(t)=\cos\left(\frac{2\,v_F\,t}{L_{so}}\right)
\end{equation}
and oscillate without any amplitude decay, as shown by the blue
dash-dotted curve in Fig. 3.

\subsection{2D ballistic channels}

Now we suppose that the channel is a 2D thin ballistic wire where
all electron spins are initially aligned in the $z$-direction, as in
the 1D case. Given an observation point, say $p_0(x_0,\,y_0)$ in
Fig. 4, the polarization $P_z(t)$ at $x_0$ at time $t$ is the
average of the spins of all electrons which will arrive at this
moment at the $x_0$ cross section. These electrons can arrive
through a straight trajectory $\overline{p_1p_0}$ of length $l=v_F
t$, or through different zigzag trajectories of the same length, as
the path $\widetilde{p_2p_0}$ in Fig. 4.

\begin{figure}[htbp!]
\center{\includegraphics[width=15cm]{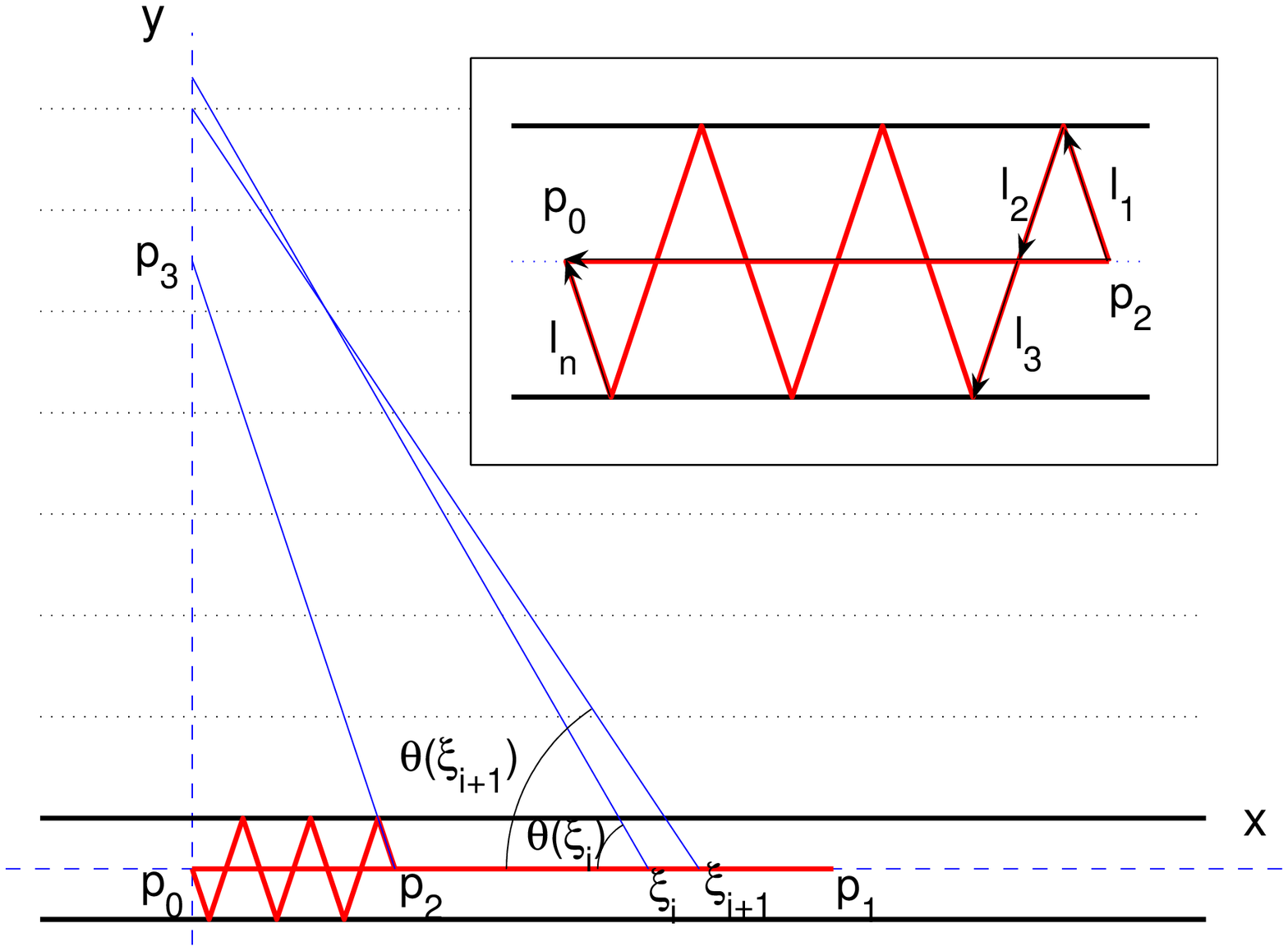}}
\caption{
(Color online) Two kinds of trajectories with the same length $l=v_F
t$ in a thin channel: the straight trajectory $\overline{p_1 p_0}$
and the zigzag trajectory $\widetilde{p_2 p_0}$. The latter has the
same length as the straight line $\overline{p_2 p_3}$, since it can
be obtained through a multiple mirror reflection of $\overline{p_2
p_3}$ with respect to the horizontal dotted lines. The symbol
$\theta(\xi)$ denotes a certain outgoing angle of an electron
located at $\xi$, as explained in the text. The trajectory
$\widetilde{p_2 p_0}$ is magnified in the inset.}
\end{figure}

As follows from Eq. (\ref{eq2_3}), the spin state of an electron
running along the zigzag trajectory $\widetilde{p_2p_0}$ in the
inset of Fig. 4 will evolve according to the spin evolution operator
\begin{eqnarray}\label{eq4_2}
   &&\hspace{-0.4cm}U_{\widetilde{p_2 p_0}}=\exp\left(\!-i\,\frac{\bm{l}_n\!\cdot\!\bm{\sigma}}{L_{so}}\right)...
   \exp\left(\!-i\,\frac{\bm{l}_2\!\cdot\!\bm{\sigma}}{L_{so}}\right)
   \exp\left(\!-i\,\frac{\bm{l}_1\!\cdot\!\bm{\sigma}}{L_{so}}\right) \nonumber \\
   &&\hspace{-0.4cm}=\left[\bm{1}-i\,\frac{\bm{l}_n\!\cdot\!\bm{\sigma}}{L_{so}}+\!...\right]\!...\!
   \left[\bm{1}-i\frac{\bm{l}_2\!\cdot\!\bm{\sigma}}{L_{so}}+\!...\right]\!
   \left[\bm{1}-i\,\frac{\bm{l}_1\!\cdot\!\bm{\sigma}}{L_{so}}+\!...\right]  \nonumber \\
   &&\hspace{-0.4cm}=U_{\overline{p_2
   p_0}}+O\left(\frac{w^2}{L_{so}^2}\right),
\end{eqnarray}
Making the above expansion up to the linear in $w$ term, the
operator $U_{\overline{p_2p_0}}$ can be written as
$$
   U_{\overline{p_2p_0}}:=\exp\left(-i\,\frac{\bm{l}\cdot\bm{\sigma}}{L_{so}}\right)
   =\left[\bm{1}-i\,\frac{\sum_{j=1}^n\bm{l}_j\cdot\bm{\sigma}}{L_{so}}+...\right],
$$
with the vector ${\bm l}$ pointing from $p_2$ to $p_0$. Equation
(\ref{eq4_2}) indicates that the spin evolution of an electron
moving along the zigzag trajectory $\widetilde{p_2 p_0}$ is
approximately the same as that of an electron drifting along the
shorter straight line $\overline{p_2p_0}$ with a drift velocity $v_F
x/l$ slower than $v_F$, where
$l=|\bm{l}_1|+|\bm{l}_2|+...+|\bm{l}_n|$ and
$x=|\bm{l}_1+\bm{l}_2+...+\bm{l}_n|$. As discussed in the previous
subsection, if an electron moves a distance $x$, its spin will
rotate the angle $2x/L_{so}$ in the spin space. If initially this
spin is aligned along the z-direction, its z-component will become
\begin{equation}\label{eq4_XX}
   s_z(x)=\cos\left(\frac{2x}{L_{so}}\right)+O\left(\frac{w^2}{L_{so}^2}\right).
\end{equation}

To determine how many electrons will contribute to $P_z(t)$, let us
uniformly divide the channel axis into small intervals
$[\xi_i,\,\xi_{i+1}]$ of length $\varepsilon$ separated by points
$\xi_i$ with $i=0,\,1,\,2,\,...$. Let $\theta(\xi_i)$ be the
outgoing angle of an electron at $\xi_i$. This angle is chosen so
that when the electron travel a zigzag path of the length $l=v_F t$,
its drift length will be $x$. Hence, this electron will arrive at
$p_0$ at time $t$. If outgoing angles are isotropically distributed,
the number of such electrons within $[\xi_i,\,\xi_{i+1}]$ is
proportional to the spanned angle
$W(\xi_i)=\theta(\xi_i)-\theta(\xi_{i+1})$. For a given small
interval $\varepsilon=\xi_{i+1}-\xi_i$ this angle is related to $x$
by
\begin{eqnarray}\label{eq4_3}
W(x) \! &=& \!\theta(x)-\theta(x+\varepsilon)
   = \arccos\left(\frac{x}{l}\right)\!-\arccos\left(\frac{x+\varepsilon}{l}\right) \nonumber \\
&=&\frac{\varepsilon}{\sqrt{l^2-x^2}}+O\left(\frac{\varepsilon}{l}\right),
\end{eqnarray}
where the outgoing angle $\theta(x)$ of the trajectory along
$\widetilde{p_2p_0}$ is the same as the angle of $\overline{p_2p_3}$
(see Fig. 4). According to Eq. (\ref{eq2_5}), the spin polarization
$P_z(t)$ at $p_0$ will be contributed from electrons traveling from
different initial locations $x$:
\begin{equation}\label{eq4_4}
    P_z(t)=\frac{\displaystyle \int_0^l \rho\,W(x)\,s_z(x)\,dx}
    {\displaystyle \int_0^l \rho\,W(x)\,dx},
\end{equation}
where $\rho$ is the line density of electrons along the channel
axis. Inserting Eq. (\ref{eq4_XX}) and (\ref{eq4_3}) into Eq.
(\ref{eq4_4}) yields
\begin{eqnarray*}\label{}
    P_z(t)&=& \frac{\displaystyle\int_0^l\frac{\varepsilon}{\sqrt{l^2-x^2}}\,
    \cos\left(\frac{2x}{L_{so}}\right)dx}{\displaystyle\int_0^l\frac{\varepsilon}{\sqrt{l^2-x^2}}\;dx}
    \\
    &=& J_0\left(\frac{2v_F\,t}{L_{so}}\right)+O\left(\frac{w^2}{L_{so}^2}\right),
\end{eqnarray*}
which at the $w\rightarrow 0$ limit is the Bessel function in Eq.
(\ref{eq3_2}). Hence, uniformly polarized spins in a ballistic
narrow channel will relax to the zero polarization through a Bessel
function. This phenomenon is in contrast to our conventional
intuition that a relaxation is a monotonically exponential process.
It is worthwhile to note that the Bessel-like spin dynamics also
takes place in other SOI systems.\cite{Culcer}

We conclude that the spin relaxation dynamics even in a very thin 2D
channel is remarkably different from that in a 1D channel, as can be
seen from a comparison of Eq. (\ref{eq3_2}) with Eq. (\ref{eq4_1}).
It can be understood from the fact that no matter how narrow the
width of a 2D channel is, it contains a large number of electrons
moving along various zigzag trajectories bouncing between two
channel boundaries. These trajectories give a significant
contribution and change the dynamics of $P_z(t)$ from a sinusoidal
oscillation in 1D systems to a Bessel-function decay in 2D systems.

\section{V Relaxation of helix spin modes}

In previous sections, all initial electron spins were polarized
along the $z$-axis. The relaxation dynamics of such spin
configuration does not change dramatically at small $w$ (the maximum
$\tau_s$ only reaches 28 ps in Fig. 1.), in agreement with the
experiment.\cite{Awschalom}. In fact, this behavior is expected from
the analysis of eigenstates of the spin diffusion equation. As shown
in Ref. \onlinecite{Malshukov}, the only homogeneous eigenstate is
$\Psi_0$ in Fig. 5(a), which has all spins polarized in the
$y$-direction and whose relaxation time strongly increases in narrow
wires. In addition to this mode there are two nonuniform slowly
relaxing eigenstates.

\begin{figure}[htbp!]
\center{\includegraphics[width=15cm]{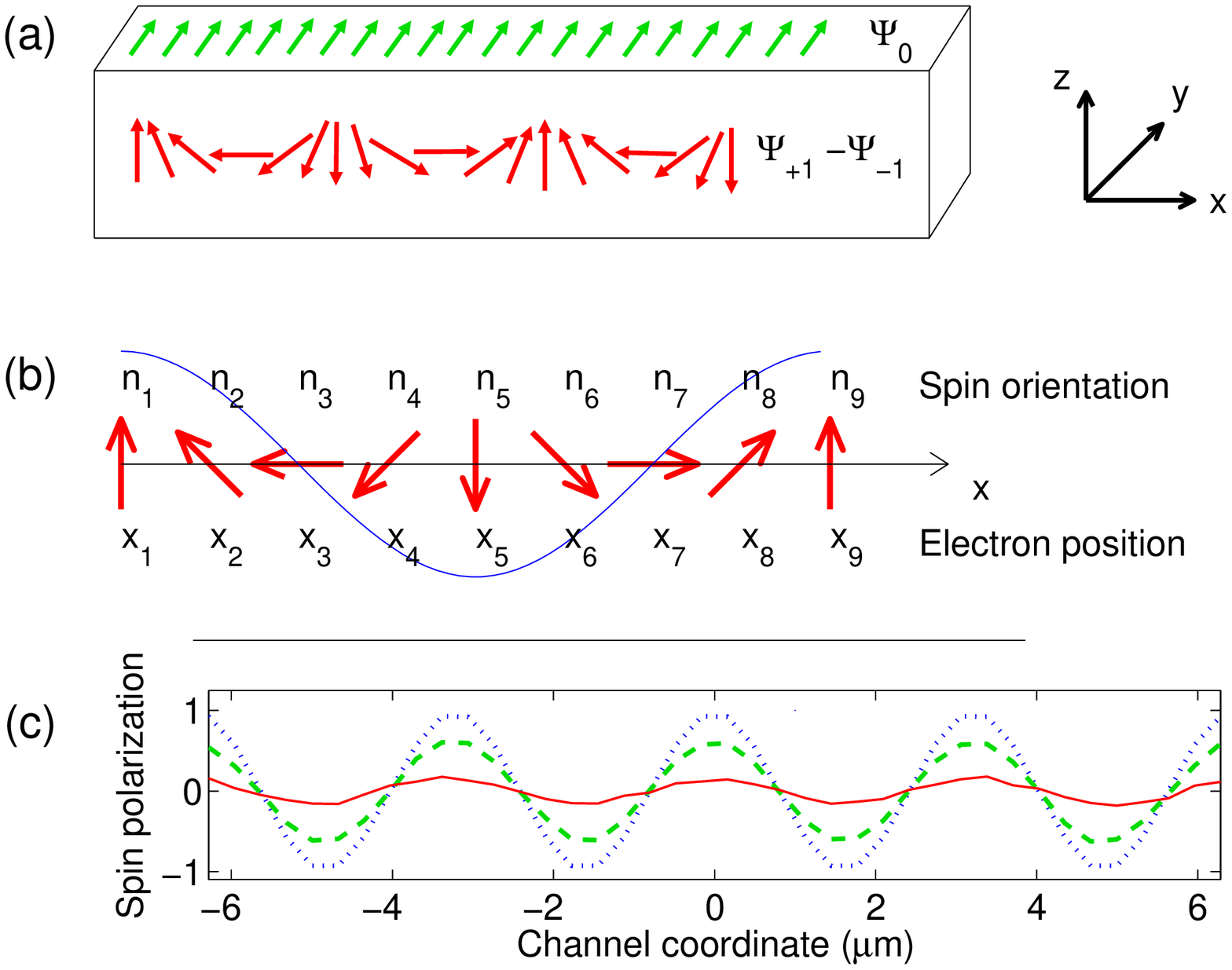}}
\caption{
(Color online) (a) Long-lived spin eigenmodes: $\Psi_{0}$ denotes
the spin mode with all spins aligned in $y$-direction.
$\Psi_{+1}-\Psi_{-1}$ represents the helix spin modes with spin
rotating on the xz-plane. (b) An electron in the channel moves a
distance $\pi L_{so}$ from the left end $x_{1}$ to the right end
$x_{9}$. Due to Rashba SOI, the spin of the electron precesses from
$\bm{n}_{1}$ to $\bm{n}_{9}$ and completes a phase period of $2\pi$.
(c) A schematic plot of the decay of the helix mode at $1$ (dotted),
$35$ (dash), and $170$ (solid) time units.
$\Psi_{\pm 1}$ in Fig. 5(a) has been replaced by
$\Psi_{+1}-\Psi_{-1}$.}
\end{figure}

These two long-lived eigenmodes exist in a 2D channel where $w\ll
l_{SP}$. Using a perturbation method with respect to $w/l_{SP}$, one
can solve a diffusion equation and obtain its unperturbed
eigensolution \cite{Malshukov}
\begin{equation}\label{eq5_1}
\widetilde{\psi}_{M,k,m}(x,y)=\exp(ikx)\chi_{m}(y)\Psi_{M},
\end{equation}
with the eigenvalue (relaxation rate)
\begin{equation}\label{eq5_2}
\Gamma^{0}_{M,k,m}=D(\pi m/d)^2+D(k-Ml_{SP}^{-1})^{2}.
\end{equation}
Therein, $\Psi_{M}$ are the eigensolutions of the momentum operator
$J_{y}$ with eigenvalues $M=0, \pm 1$ and $\chi_{2n}(y)=\cos(2\pi
yn/d)$, as well as $\chi_{2n+1}(y)=-\sin[\pi y(2n+1)/d]$. In Fig.
5(a), $\Psi_{\pm 1}$ will construct the helix eigenmodes with the
wave vectors $k=\mp 1/l_{SP}$. We note that the eigenmode of the
diffusion equation is a spin configuration exponentially decreasing
in time, but its shape remaining unchanged. Taking the second order
correction in $w/l_{SP}$, one obtains
\begin{subequations}
\begin{eqnarray}
\Gamma_{M,k,m}&=&\Gamma^{0}_{M,k,m}+\frac{(2-M^{2})w^{2}}{24\tau_{s0}\;l_{SP}^2} \label{eq5_3} \\
&=&\Gamma^{0}_{M,k,m} +\frac{2(2-M^{2}) w^2 v_F l}{3L_{so}^{4}}
\label{eq5_4}
\end{eqnarray}
\end{subequations}
for $\mid k-M/l_{SP}\mid \ll 1/l_{SP}$, where
$\tau_{s0}=l_{SP}^2/v_F l$ and $l_{SP}=L_{so}/2$ have been used in
the second equality. The modes with $m=0$, and $k=M/l_{SP}$ will
relax most slowly, since in these cases the first term
$\Gamma^{0}_{M,k,m}$ in Eq. (15) disappears. The second term
indicates that the spin relaxation time $\tau_{s}=1/\Gamma_{M,k,m}$
will be proportional to $1/w^2$. In the limit $L_{so}\gg w$, we thus
have $\tau_{s}$ much larger than the D'yakonov Perel' relaxation
time $\tau_{s0}$ in 2D boundless systems.\cite{DP} Such a behavior
will become more clear from the following simple consideration in a
1D system. Let us consider an electron at $x_1$ in a 1D channel with
an initial spin pointing to $\bm{n}_1$, as shown in Fig. 5(b). Under
the Rashba SOI, the spin will rotate to $\bm{n}_2$, $\bm{n}_3$, ...,
when this electron moves to $x_2$, $x_3$, .... It is easy to see
that if each electron spin in an initial spin density distribution
follows this ($x_i,\,\bm{n}_i$) relation, such a distribution will
not change in time. Hence, its relaxation time is infinite. In a
realistic 2D wire the relaxation time is finite at finite width.
That is because electrons there can move in the y-direction. The
polarization $P_z(t)$ contributed from electrons moving along the
channel is frozen, as in the 1D case, while electrons moving along
the the y-axis give rise to the relaxation of the helix
distribution.

Since the above expressions have been obtained under the assumption
$l\ll w\ll L_{so}$, it is interesting to extend the analysis beyond
this limits. Within the SPI method we studied the relaxation of the
helix mode in the range $l \lesssim w$, choosing
$L_{so}=12.5\,\mu$m, $l=0.5\,\mu$m, and $v_F=0.37\mu$m/ps. Given the
initial helix spin mode $\Psi_{+1}-\Psi_{-1}$ corresponding to spins
oriented along the z-axis at $x$=0 in Fig. 5(a), the spin relaxation
time $\tau_s(w)$ calculated by the SPI method is represented by
squares in the inset of Fig. 6. In the range of $w<10\,\mu$m, this
time increases dramatically and strongly deviates from the $\tau_s$
of the corresponding uniform mode (red solid curve). Note that in
order to display the divergent $\tau_s$, we choose a large
$\tau_s$-axis scale in the inset of Fig. 6. At this scale the
uniform mode $\tau_s$ (red curve) almost overlaps with the
$\tau_s=0$ axis. The relaxation time of the uniform mode will
saturate at some value for $w\rightarrow 0$. It is $\tau_s\approx
28$ ps for $[L_{so},\,l,\,w]=[2.19,\,0.28,\,1.4]\,\mu$m in Fig. 1
and $\tau_s\approx 441.1$ ps for
$[L_{so},\,l,\,w]=[12.5,\,0.5,\,0.4]\,\mu$m. In contrast to these
uniform modes, $\tau_s$ of the helix mode strongly increases for
$w\rightarrow 0$ (if $w\gg l$). This behavior is consistent with the
theoretical result Eq. (\ref{eq5_3}). On the other hand, if $w$ is
as large as $20\,\mu$m, the relaxation time $\approx 1927.4$ ps of
the helix mode is still much larger than $\tau_s\approx 241.0$ ps
corresponding to the uniform mode. This difference can be easily
seen by magnifying the $\tau_s$-axis of the inset in Fig. 6.

\begin{figure}[htbp!]
\center{\includegraphics[width=15cm]{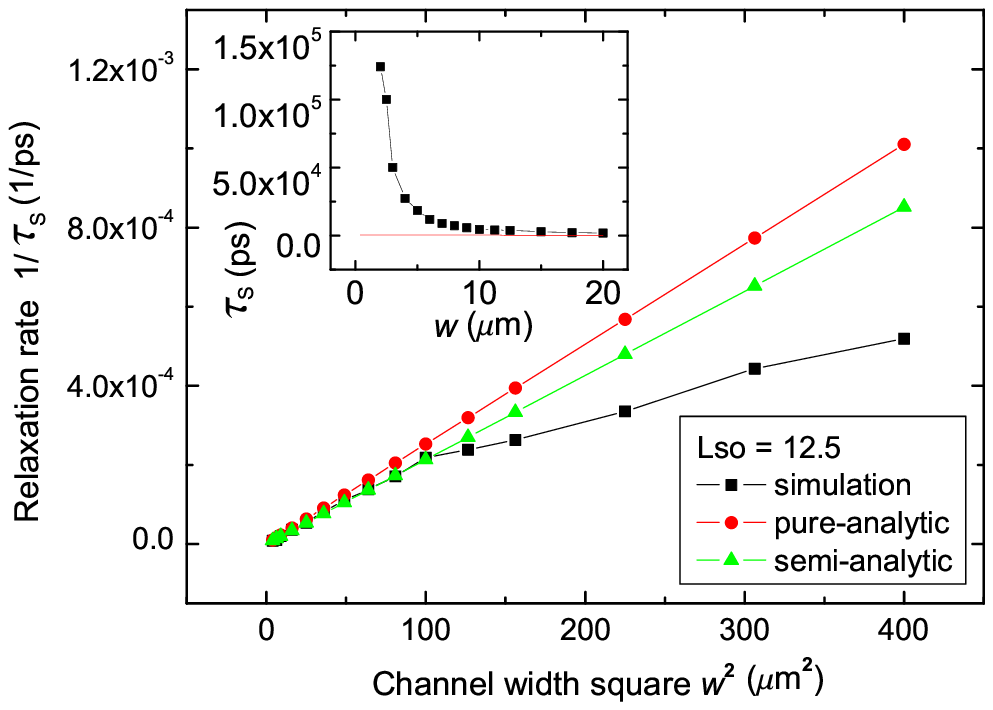}}
\caption{
(Color online) The relaxation time $\tau_s$ of the uniform mode (red
solid) and the helix mode (black square) versus the channel width
$w$ in the inset. For small $w$, the square curve of the helix mode
is redrawn as $1/\tau_s$ against $w^2$ in the main plot, to compare
with the $\tau_s$'s derived from the analytical formula
(\ref{eq5_4}) (red circle) and the semi-analytical formula
(\ref{eq5_3}) (green triangular).}
\end{figure}

The main plot in Fig. 6 shows the  $1/\tau_s$ dependence on $w^2$.
The squares show the helix relaxation rate corresponding to the data
in the inset. The line with circles for the helix mode is obtained
from the analytical expression Eq. (\ref{eq5_4}), while the line
with triangles is obtained from Eq. (\ref{eq5_3}). In the latter
line, $\tau_{s0}$ is simulated numerically, instead of using the
analytical relation $\tau_{s0}=l_{SP}^2/v_F l$ mentioned below Eq.
(\ref{eq5_4}). The lines with circles and triangles are valid only
at $l\ll w \ll L_{so}$, while that with squares is valid for all
$w$. One can clearly see that the curve calculated by the SPI method
agrees very well with Eq. (15) at small $w$ between $w_1=2\,\mu$m
(the smallest calculated width) and $w_2=10\,\mu$m. It is
interesting to note that although formula Eq. (15) was derived under
the condition $l\ll w\ll L_{so}$, it seems to be valid in a wider
range of $w$, because $w_1$ is close to the ballistic regime
crossover point at $w \sim l=0.5\,\mu$m and $w_2$ is close to
$L_{so}=12.5\,\mu$m. Finally, one should not expect that the linear
trend in Fig. 6 can continue down to $w^2\rightarrow 0$ because in
this range the system will eventually reach the ballistic regime and
$P_z(t)$ will decay like a crossover function between the
exponential and the Bessel functions. Similar to the discussion of
the uniform initial spin configuration in Section III, it does not
make sense to consider $\tau_s$ at extremely small $w$, because
$P_z(t)$ is no longer an exponential function.

\section{VI Conclusion}

The semiclassical path integral method has been applied to study the
spin relaxation in thin 2D wires with the Rashba spin-orbit
interaction. We considered the relaxation of a uniform spin
polarization along the $z$-axis, as well as of the long-lived helix
mode. In the former case we found a good agreement of $\tau_s$
calculated in the regime of large $w$ ($w\approx 20\,\mu$m) with the
well known bulk DP spin relaxation rate and with the experimental
data from Ref. \onlinecite{Awschalom}. At smaller $w$ our numerical
results deviate slightly from the experimental data. The nature of
this distinction is not clear. We assume that the edge spin
diffusion modes can contribute to the spin relaxation, so that the
conditions for electron reflections from the wire lateral boundaries
become important. Also, the Dresselhaus spin-orbit interaction can
give rise to the observed dependence of $\tau_s$ on the orientation
of the wire axis in the xy-plane. At $w \rightarrow 0$ the
relaxation time has a tendency to saturate at a value which is about
twice of the bulk $\tau_s$, as predicted by the spin diffusion
theory. Although both SPI and experimental data show similar
saturation behavior, one must take into account that at $w$
approaching the crossover $w \sim l_{f}$ to the ballistic regime the
evolution of the spin polarization can not be described by an
exponential function. Hence, in this regime $\tau_s$ is not a
representative parameter to describe the spin relaxation. We studied
the evolution of the spin polarization in the ballistic regime and
found that it is described by the Bessel function. The numerical SPI
results fit well to this behavior. For the helix spin distribution,
the linear dependence of $1/\tau_s$ on $w^2$ predicted in the
framework of the diffusion theory \cite{Malshukov} coincides
precisely with that calculated by the SPI method. The SPI method
also allowed to calculate the spin relaxation beyond the constraints
of those analytic results. \cite{Malshukov}

This work is supported by the National Science Council in Taiwan
through Grant No. NSC 96-2112-M-009-003 and Russian RFBR Grant, No
060216699.

\end{document}